\documentclass[aps,prl,twocolumn,superscriptaddress]{revtex4}
\usepackage{times}
\usepackage{amsmath}
\usepackage{textcomp}
\usepackage[pdftex]{graphicx}
\usepackage{float}
\usepackage{verbatim}
\usepackage{color}
\definecolor{darkgreen}{rgb}{0,0.6,0}
\definecolor{purple}{rgb}{0.63, 0.36, 0.94}
\makeatletter
\renewenvironment{widetext@grid}{%
  \par\ignorespaces
  \setbox\widetext@top\vbox{%
   \vskip15\p@
   \hb@xt@\hsize{%
    \leaders\hrule\hfil
    \vrule\@height6\p@
   }%
   \vskip6\p@
  }%
  \setbox\widetext@bot\hb@xt@\hsize{%
    \vrule\@depth6\p@
    \leaders\hrule\hfil
  }%
  \onecolumngrid
  \let\set@footnotewidth\set@footnotewidth@ii
}{%
  \par
  \twocolumngrid\global\@ignoretrue
  \@endpetrue
}%
\makeatother

\begin{document}

\title{A Two Qubit Logic Gate in Silicon}

\author{M. Veldhorst}
\affiliation{Centre for Quantum Computation and Communication Technology, School of Electrical Engineering and Telecommunications, The University of New South Wales, Sydney, NSW 2052, Australia}
\author{C.H. Yang}
\affiliation{Centre for Quantum Computation and Communication Technology, School of Electrical Engineering and Telecommunications, The University of New South Wales, Sydney, NSW 2052, Australia}
\author{J.C.C. Hwang}
\affiliation{Centre for Quantum Computation and Communication Technology, School of Electrical Engineering and Telecommunications, The University of New South Wales, Sydney, NSW 2052, Australia}
\author{W. Huang}
\affiliation{Centre for Quantum Computation and Communication Technology, School of Electrical Engineering and Telecommunications, The University of New South Wales, Sydney, NSW 2052, Australia}
\author{J.P. Dehollain}
\affiliation{Centre for Quantum Computation and Communication Technology, School of Electrical Engineering and Telecommunications, The University of New South Wales, Sydney, NSW 2052, Australia}
\author{J.T. Muhonen}
\affiliation{Centre for Quantum Computation and Communication Technology, School of Electrical Engineering and Telecommunications, The University of New South Wales, Sydney, NSW 2052, Australia}
\author{S. Simmons}
\affiliation{Centre for Quantum Computation and Communication Technology, School of Electrical Engineering and Telecommunications, The University of New South Wales, Sydney, NSW 2052, Australia}
\author{A. Laucht}
\affiliation{Centre for Quantum Computation and Communication Technology, School of Electrical Engineering and Telecommunications, The University of New South Wales, Sydney, NSW 2052, Australia}
\author{F.E. Hudson}
\affiliation{Centre for Quantum Computation and Communication Technology, School of Electrical Engineering and Telecommunications, The University of New South Wales, Sydney, NSW 2052, Australia}
\author{K.M. Itoh}
\affiliation{School of Fundamental Science and Technology, Keio University, 3-14-1 Hiyoshi, Kohoku-ku, Yokohama 223-8522, Japan.}
\author{A. Morello}
\affiliation{Centre for Quantum Computation and Communication Technology, School of Electrical Engineering and Telecommunications, The University of New South Wales, Sydney, NSW 2052, Australia}
\author{A.S. Dzurak}
\affiliation{Centre for Quantum Computation and Communication Technology, School of Electrical Engineering and Telecommunications, The University of New South Wales, Sydney, NSW 2052, Australia}

\date{\today}

\begin{abstract}
Quantum computation requires qubits that can be coupled and realized in a scalable manner, together with universal and high-fidelity one- and two-qubit logic gates \cite{DiVincenzo2000, Loss1998}. 
Strong effort across several fields have led to an impressive array of qubit realizations, including trapped ions \cite{Brown2011}, superconducting circuits \cite{Barends2014}, single photons\cite{Kok2007}, single defects or atoms in diamond \cite{Waldherr2014, Dolde2014} and silicon \cite{Muhonen2014}, and semiconductor quantum dots \cite{Veldhorst2014}, all with single qubit fidelities exceeding the stringent thresholds required for fault-tolerant quantum computing \cite{Fowler2012}. Despite this, high-fidelity two-qubit gates in the solid-state that can be manufactured using standard lithographic techniques have so far been limited to superconducting qubits \cite{Barends2014}, as semiconductor systems have suffered from difficulties in coupling qubits and dephasing \cite{Nowack2011, Brunner2011, Shulman2012}. Here, we show that these issues can be eliminated altogether using single spins in isotopically enriched silicon\cite{Itoh2014} by demonstrating single- and two-qubit operations in a quantum dot system using the exchange interaction, as envisaged in the original Loss-DiVincenzo proposal \cite{Loss1998}. We realize CNOT gates via either controlled rotation (CROT) or controlled phase (CZ) operations combined with single-qubit operations. Direct gate-voltage control provides single-qubit addressability, together with a switchable exchange interaction that is employed in the two-qubit CZ gate. The speed of the two-qubit CZ operations is controlled electrically via the detuning energy and we find that over 100 two-qubit gates can be performed within a two-qubit coherence time of 8 \textmu s, thereby satisfying the criteria required for scalable quantum computation.
 
\end{abstract}

\maketitle
Quantum dots have high potential as a qubit platform \cite{Loss1998}. Large arrays can be conveniently realized using conventional lithographic approaches, while reading, initializing, controlling and coupling can be done purely by electrical means. Early research focussed mainly on III-V semiconductor compounds such as GaAs, resulting in single spin qubits \cite{Koppens2006}, singlet-triplet qubits \cite{Petta2005} and exchange only qubits \cite{Medford2013}, which can be coupled capacitively \cite{Shulman2012} or via the exchange interaction \cite{Nowack2011, Brunner2011}. While these approaches demonstrate the potential of quantum dot qubits, strong dephasing due to the nuclear spin background have limited the quality of the quantum operations. A strong improvement in coherence times has been observed by defining the quantum dots in silicon \cite{Maune2012, Kawakami2014}, which can be isotopically purified \cite{Itoh2014}, such that quantum dots with single spin fidelities above the threshold of surface codes \cite{Fowler2012} can be realized \cite{Veldhorst2014}.

A scalable approach towards quantum computation ideally requires that the coupling between qubits can be turned on and off \cite{DiVincenzo2000}, so that single and two-qubit operations can be selectively chosen. Here, we demonstrate this by realizing a CZ gate, which is commonly used in superconducting qubits \cite{Barends2014} and has been theoretically discussed for quantum dot systems \cite{Meunier2011}. This two-qubit gate, together with single-qubit gates provides all of the necessary operations for universal quantum computation. We can control qubits individually by tuning the qubit resonance frequency via electrical gate-voltage control. When the coupling is turned off, individual qubit operations are performed using a global electron-spin-resonance (ESR) line and the qubits can rely on the long coherence times provided by the nuclear spin-free background \cite{Veldhorst2014}. When the coupling is turned on, conditional operations such as the CNOT can be performed, incorporating a CZ gate with operation time 100 ns. From direct measurement of the ESR transitions of the two-qubit system we are also able to fully map out the exchange coupling as a function of detuning. 

\begin{figure*} [t!]
	\centering 
		\includegraphics[width=180mm]{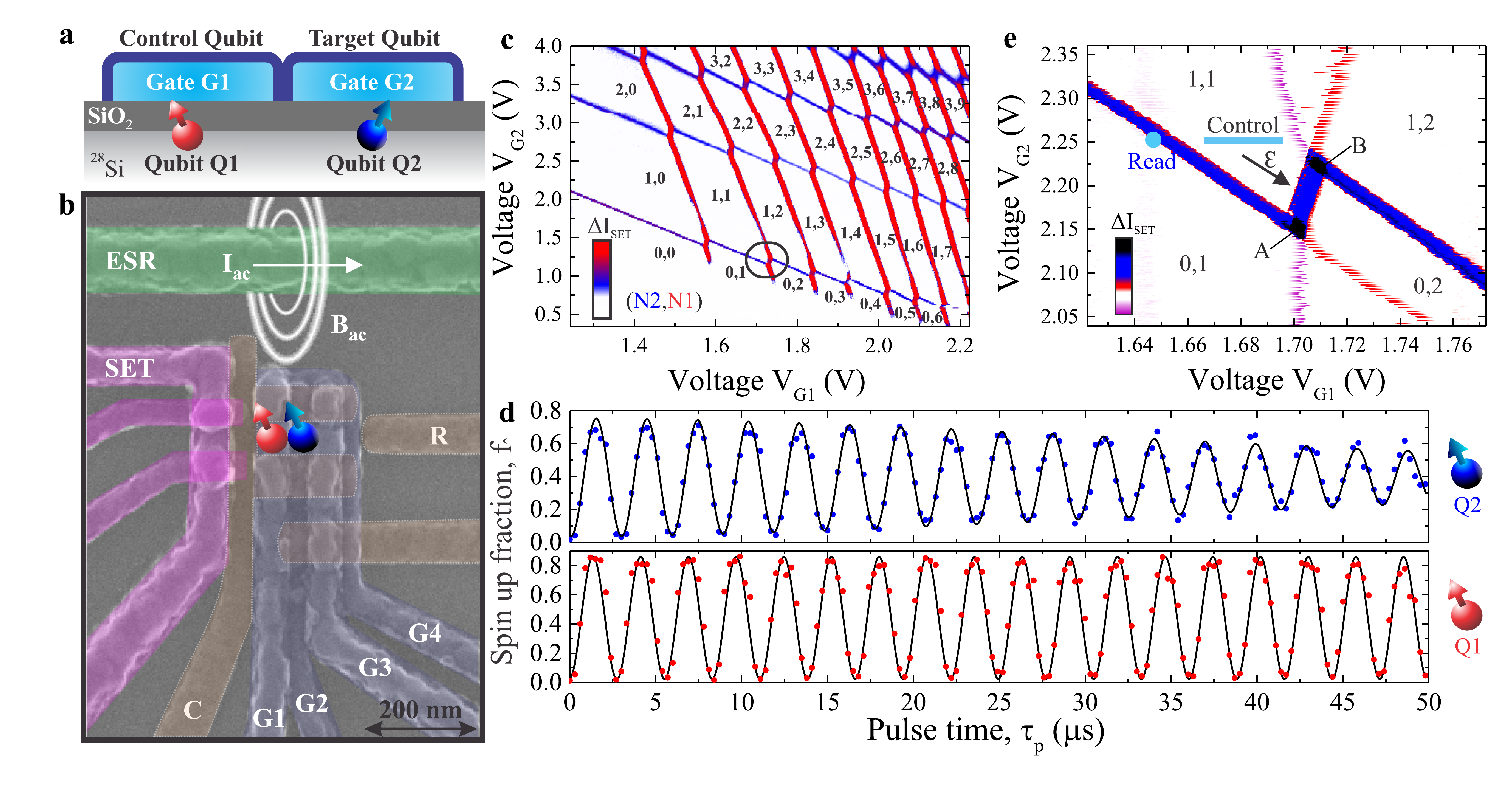}
		\caption{ \textbf{Silicon two-qubit logic device, incorporating SET readout and selective qubit control. a} Schematic and \textbf{b} SEM coloured image of the device. The quantum dot structure (labels C and G) can be operated as a single or double quantum dot by appropriate biasing of gate electrodes G1-G4, where we choose here to confine the dots underneath G1 and G2.  The confinement gate C runs underneath the gates G1-G3 and confines the quantum dot on all sides except on the reservoir side (R). \textbf{c} Stability diagram of the double quantum dot obtained by monitoring the capacitively coupled SET. The difference in distance towards the SET results in different capacitive coupling, such that the individual dots can be easily distinguished. Interestingly, we see that the tunnel coupling of the fourth transition ($\textrm{N}_1$ = 3$\rightarrow$4) of dot 1 is relatively weak, which is due to valley and spin filling, since there is only one state in the lowest orbital that can be occupied. Qubit $\textrm{Q}_1$ and qubit $\textrm{Q}_2$ are realized by depleting dot 1 and dot 2 to the last electron. \textbf{d} The quantum dot qubits can be individually controlled by electrically tuning the ESR resonance frequency using the Stark shift \cite{Veldhorst2014}. Clear Rabi oscillations for both qubits are observed.	\textbf{e} Zoom-in on the $(1,1)$-$(0,2)$ charge states (circled in \textbf{c}); the region where we operate the two-qubit system. Tunnelling of dot 1 to the reservoir R is via dot 2 and results in a hysteresis as function of the sweeping direction due to finite mutual charging energy \cite{Yang2014}. All measurements were performed in a dilution refrigerator with base temperature $T \approx$ 50 mK and a dc magnetic field $B_0$ = 1.4 T.}  
		\label{fig:1}
\end{figure*}

Figure \ref{fig:1}a shows a schematic and Fig. \ref{fig:1}b shows an SEM-image of the double quantum dot structure fabricated on a $^{28}$Si epilayer with a residual $^{29}$Si concentration of 800 ppm \cite{Fukatsu2003}. The device consists of three aluminum layers, nine aluminum gates, an aluminum lead for ESR control \cite{Dehollain2013}, and source, drain and reservoir leads that connect to the gate-induced two dimensional electron gas (2-DEG) using multi-level gate-stack silicon MOS technology \cite{Angus2007}. A single-electron-transistor (SET) is formed to monitor the charge state of the quantum dot system and for spin read out using spin-to-charge conversion \cite{Elzerman2004}. 

Figure \ref{fig:1}c shows the stability diagram of the double quantum dot system with charge occupancy ($\textrm{N}_1$,$\textrm{N}_2$). The transitions of quantum dot 1, which is underneath gate G1, and quantum dot 2, which is underneath gate G2, can be distinguished by their gate voltage dependence and their capacitive coupling to the SET. When the quantum dot system is depleted to the few-electron regime, the tunnel coupling decreases, partly due to the decrease of the dot size. When dot 2 is empty, the coupling between the reservoir and dot 1 vanishes and the transitions of dot 1 disappear. We define qubit $\textrm{Q}_1$ by loading a single electron into dot 1, so that $\textrm{N}_1$=1, and similarly for qubit $\textrm{Q}_2$ we have $\textrm{N}_2$=1. 

We operate the qubits individually by applying an ac magnetic field $B_{ac}$ at their resonance frequency via the ESR line \cite{Veldhorst2014}. Clear Rabi oscillations are observed for both qubits, as shown in Fig. \ref{fig:1}d. The visibility of qubit $\textrm{Q}_1$ is slightly higher than that of qubit $\textrm{Q}_2$ due to the stronger capacitive coupling to the SET. Qubit $\textrm{Q}_1$ has a coherence time of $T_2^*$ = 120 \textmu s, with a $T_2$ that can be extended up to 28 ms using CPMG pulses \cite{Veldhorst2014}. Qubit $\textrm{Q}_2$ has a slightly shorter coherence time and using a Ramsey experiment we find $T_2^*$ = 61 \textmu s, comparable to its Rabi decay time (see Supplementary Information section 2). We ascribe this difference to finite coupling of the reservoir to qubit $\textrm{Q}_2$, since we observed that the coherence time increased with decreasing reservoir coupling, achieved by biasing the qubit states further away from the Fermi level in the reservoir. 

Spin qubits can be coupled capacitively, as proposed by Taylor $et$ $al.$ \cite{Taylor2005}, but here we couple the qubits directly via the exchange interaction as discussed in the seminal work of Loss and DiVincenzo \cite{Loss1998}, which is expected to be the mechanism with the faster two-qubit operations. The exchange coupling can be controlled with the detuning energy $\epsilon$ as depicted in Fig.\ref{fig:1}e, which controls the energy separation between the $(1,1)$ and $(0,2)$ charge states. In Fig.\ref{fig:1}e, we have lowered the reservoir-dot 2 coupling so that the tunneling time is \texttildelow 100 \textmu s, which is the coupling during read out in the two-qubit experiments. In this range of weak reservoir-dot 1 coupling, the emptying and filling of dot 1 is hysteretic with gate voltage, since the mutual charging energy becomes relevant, as dot 1 can only tunnel when it aligns in energy with dot 2 \cite{Yang2014}. At the nodes of the $(1,1)$-$(0,2)$ crossing (A and B in Fig. \ref{fig:1}e), electrons tunnel between dot 1 and the reservoir (via dot 2) and the signal is therefore maximised, whereas in the middle of the transition tunnelling is between dot 1 and dot 2 giving rise to a smaller signal, which is comparable to tunnelling between dot 2 and the reservoir. We control the two-qubit system in the $(1,1)$ region and read out $\textrm{Q}_2$ at the $(1,1)$-$(0,1)$ transition.

\begin{figure*} [t!]
	\centering 
		\includegraphics[width=180mm]{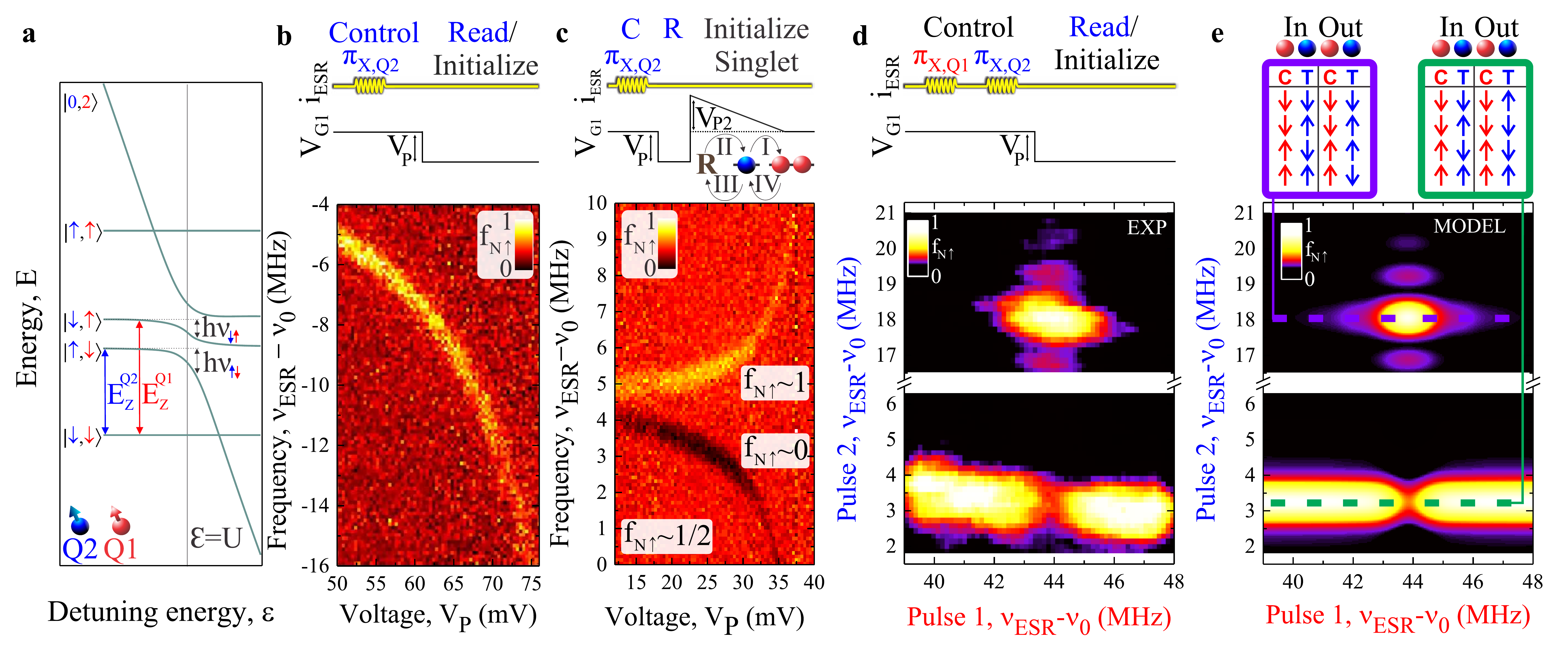}
		\caption{ \textbf{Exchange spin funnel and CNOT via CROT. a} Schematic of the coupling between qubit $\textrm{Q}_1$ and $\textrm{Q}_2$ using the exchange interaction at the $|1,1\rangle$-$|0,2\rangle$ transition. The g-factors of $\textrm{Q}_1$ and $\textrm{Q}_2$ are electrically tuned to result in a frequency difference of 40MHz (the difference is exaggerated in the schematic for clarity) for individual qubit control. \textbf{b} ESR spectrum of the $\left|{\downarrow,\downarrow} \right\rangle$-$\left|{\uparrow,\downarrow} \right\rangle$ transition as a function of increasing detuning. \textbf{c} ESR spectrum, but now with an additional pulse with amplitude $V_{P2}$ (see schematic) so that we achieve singlet initialization. Using this method we observe both the $\left|{\uparrow,\downarrow} \right\rangle$ to $\left|{\downarrow,\downarrow} \right\rangle$ and $\left|{\downarrow,\uparrow} \right\rangle$ to $\left|{\uparrow,\uparrow} \right\rangle$ transitions. Due to singlet initialization, the background spin-up fraction increases to 1/2, and the lower ESR branch results in a decay rather than an increase of the spin-up fraction. \textbf{d} Realization of a CNOT operation by applying a $\pi$-pulse to $\textrm{Q}_1$ before operation on $\textrm{Q}_2$. We find that $\textrm{Q}_2$ rotates depending on $\textrm{Q}_1$, as the $\sigma_z$-component controls the ESR resonance frequency of $\textrm{Q}_2$. \textbf{e}  We find qualitative agreement with a model that includes decoherence, where the $\left|\uparrow,\downarrow \right\rangle$-state has the shortest coherence time of \texttildelow 2 \textmu s. From \textbf{a} we see that this state has the largest $\delta \nu / \delta V_P$, and is therefore the most sensitive to electrical noise. The tables in \textbf{e} show the two-qubit spin logic operations at the purple and green dotted lines, respectively. The data of \textbf{b} and \textbf{d} have been obtained using the same voltages on all the gates, while the voltages used for the experiment shown in \textbf{c} are different. The data has been offset by a frequency $\nu_{0}$ = 39.14GHz for clarity.}  
		\label{fig:2}
\end{figure*}

The general protocol for the two-qubit operation consists of applying microwave pulses on control qubit $\textrm{Q}_1$ and target qubit $\textrm{Q}_2$ and voltage pulses to tune the exchange coupling between the qubits, followed by readout on $\textrm{Q}_2$. By exploiting the Stark shift \cite{Veldhorst2014}, we electrically control the effective $g$-factor of $\textrm{Q}_1$ and $\textrm{Q}_2$, to tune the Zeeman energy $E_Z^{1,2}=g_{1,2} \mu_B B_Z$ and the associated qubit resonance frequency $\nu_{1,2}=E_Z^{1,2}/h$ in order to realise individual qubit control. While the spin-orbit coupling in silicon is small \cite{Tahan2005}, the valley degeneracy in silicon can complicate the qubit coupling \cite{Culcer2010} and even the single-qubit operation \cite{Kawakami2014}. However, the presence of a large valley splitting energy in silicon MOS quantum dots \cite{Yang2013, Veldhorst2014}, combined with a large on-site Coulomb energy $U$ compared to the inter-dot tunnel coupling $t_0$, allows us to neglect all two-particle processes that can occur. We can consequently describe the system in the rotating wave approximation in the basis $\Psi=[\left|\color{blue}\uparrow \color{black},\color{red}\uparrow \color{black}\right\rangle,\left|\color{blue}\uparrow \color{black},\color{red}\downarrow \color{black} \right\rangle,\left|\color{blue}\downarrow \color{black},\color{red}\uparrow \color{black}\right\rangle,\left|\color{blue}\downarrow \color{black},\color{red}\downarrow \color{black}\right\rangle,\left|\color{blue}0 \color{black},\color{red}2 \color{black}\right\rangle]$ with the effective Hamiltonian:
\begin{equation}
H=\left[
\begin{array}{ccccc}
\overline{E_z}-\omega & \color{red} \Omega &\color{blue} \Omega & 0 & 0 \\
\color{red} \Omega & \delta E_Z/2 & 0 &\color{blue} \Omega &\color{darkgreen} t_0 \\
\color{blue} \Omega & 0 & -\delta E_Z/2 &\color{red} \Omega &\color{darkgreen} -t_0\\
0 & \color{blue} \Omega & \color{red} \Omega & -\overline{E_z}+\omega & 0 \\ 
0 &\color{darkgreen} t_0 &\color{darkgreen} -t_0 & 0 & U-\epsilon \end{array} \right]
\label{eq:DQD}
\end{equation} 
where $\overline{E_z}$ is the mean Zeeman energy and $\delta E_Z$ is the difference in Zeeman energy between the dots, $\epsilon$ is the detuning energy and for simplicity we have used color coding and set $\hbar=1$. single-qubit operations are realized with Rabi frequency $\Omega$, by matching the microwave frequency $\omega$ to the resonance frequency of one of the qubits. The presence of exchange coupling between the qubits alters the Zeeman levels as shown in Fig. \ref{fig:2}a, where the finite coupling between the qubits causes an anticrossing between the $|0,2\rangle$ and $|1,1\rangle$ states. We can now experimentally map out the energy levels in the vicinity of the anticrossing, as shown in Figs. \ref{fig:2}b and c. In Fig. \ref{fig:2}b, we have initialized $\textrm{Q}_1$ and $\textrm{Q}_2$ to spin down and by applying a $\pi$-pulse to $\textrm{Q}_2$ we can map out the resonance frequency of $\textrm{Q}_2$ as function of detuning. We measure exchange couplings of more than 10 MHz, above which the $T_2^*$ becomes too short for spin-flips to occur.

\begin{figure*} [t!]
	\centering 
		\includegraphics[width=180mm]{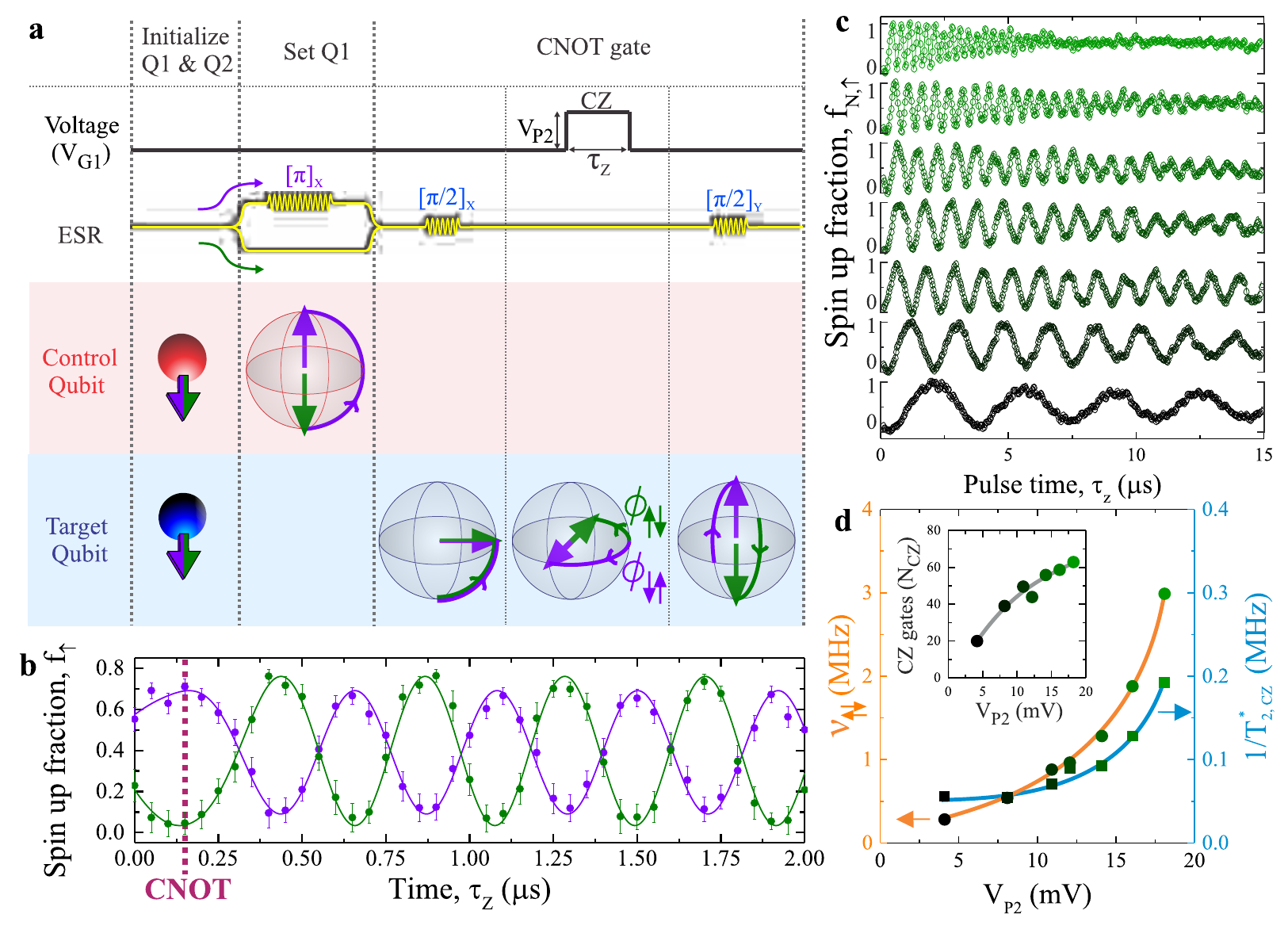}
		\caption{ \textbf{A quantum dot CZ gate. a} Pulsing protocol for CNOT operations using CZ. When the coupling is on, the qubits make in-plane rotations with a frequency determined by the exchange coupling and a direction that depend on the $\sigma_z$ of the other qubit. \textbf{b} Demonstration of the CNOT gate. The control qubit $\textrm{Q}_1$ is prepared into the $|\color{purple} {\uparrow} \color{black} \rangle$ or $|\color{darkgreen} {\downarrow} \color{black} \rangle$ state, after which a CZ($\pi$) gate is performed. Clear oscillations as a function of the interaction time $\tau_Z$ arise and are fitted by assuming a Gaussian-shaped voltage pulse due to finite fall and rise times. The error bars correspond to the standard error of the mean of 900 single shot events per data point. \textbf{c} Via the detuning energy, set with $V_{G1}$, the exchange coupling can be controlled resulting in a tunable two-qubit operation frequency $\nu_{\uparrow\downarrow}$. \textbf{d} By fitting the data of \textbf{c} we map out $\nu_{\uparrow\downarrow}$ and $T^*_{2,CZ}$ as a function of $V_{P2}$. The inset shows the number of possible CZ($\pi$) operations, where $N_{CZ}=2 \times (\nu_{\uparrow\downarrow}+\nu_{\downarrow\uparrow}) \times T^*_{2,CZ}$. While the $T^*_{2,CZ}$ decreases with coupling, the number of possible two-qubit operations continue to increase.}  
		\label{fig:3}
\end{figure*}

Initialization of the $|{\uparrow},\downarrow\rangle$-$|{\downarrow},\uparrow\rangle$ singlet state is possible by pulsing towards the $(1,2)$ and returning to the $(1,1)$ charge state. In this sequence, see Fig. \ref{fig:2}c, an electron tunnels from dot 2 to dot 1 (I) followed by an electron tunnelling from the reservoir R to dot 2 (II). After returning to (1,1), the electron from dot 2 tunnels back to R (III), and one of the two electrons on dot 1, which are in a singlet state, tunnels to dot 2 (IV). With this initialization into the singlet state, when we apply a microwave pulse on $\textrm{Q}_2$ the spin up fraction $f_\uparrow$ is always 1/2, except when the microwave frequency matches a resonance frequency of $\textrm{Q}_2$. Due to the finite exchange interaction, there are two resonance frequencies. The lower frequency rotates the singlet state towards $\left|\downarrow,\downarrow \right\rangle - \left| \downarrow,\uparrow \right\rangle$, where $\textrm{Q}_2$ always ends up as spin down. The higher frequency rotates the singlet state towards $\left|\uparrow,\downarrow \right\rangle - \left|\uparrow,\uparrow \right\rangle$, where $\textrm{Q}_2$ always ends up as spin up. The results are depicted in Fig. \ref{fig:2}c, showing a decrease in $f_\uparrow$ at the lower branch and an increase of $f_\uparrow$ at the upper branch, demonstrating an exchange spin funnel, where both branches are visible, as opposed to the single-branch spin-funnel observed in singlet-triplet qubits \cite{Petta2005}. In the Supplementary Information we show more detailed sequences of the initialization process, showing that additional levels that mix with the $\left|\downarrow,\downarrow \right\rangle$-state can be present at certain detunings. 

Having characterized the two-qubit system, we now turn to controlled operations using CROT and demonstrate a CNOT gate, see Figs. \ref{fig:2}d and e. In this sequence, the exchange interaction is always on and the resonance frequency of target qubit $\textrm{Q}_2$ depends on the $\sigma_z$-component of $\textrm{Q}_1$, so that there is a resonance frequency where $\textrm{Q}_2$ only rotates when $\textrm{Q}_1$ is spin up and a resonance frequency where $\textrm{Q}_2$ only rotates when $\textrm{Q}_1$ is down. We set the exchange coupling such that the two resonance frequencies of $\textrm{Q}_2$ are split by 14.5 MHz and apply a $\pi$-pulse on the control qubit $\textrm{Q}_1$, followed by a $\pi$-pulse on the target qubit $\textrm{Q}_2$. The experimental results shown in Fig. \ref{fig:2}d coincide reasonably with the model shown in Fig. \ref{fig:2}e, which includes decoherence. The side lobes of the $\pi$-pulse can be observed for the $\left|\downarrow,\uparrow \right\rangle$ to $\left|\uparrow,\uparrow \right\rangle$ transition, but not for the $\left|\downarrow,\downarrow \right\rangle$  to $\left|\uparrow,\downarrow \right\rangle$ transition. We ascribe this difference to the $g$-factor difference between the qubits, which gives a difference in coupling between the $\left|\uparrow,\downarrow \right\rangle$ and $\left|\downarrow,\uparrow \right\rangle$  to the $\left|0,2 \right\rangle$-state, see Eq. 1, resulting in a different sensitivity to electrical noise.

The presence of exchange coupling makes the system more vulnerable to electrical noise sources such as electrical gate noise, thereby lowering the coherence time and limiting the amount of possible operations using CROT. However, when the coupling is off, single-qubit fidelities above some fault-tolerant thresholds are possible \cite{Veldhorst2014}. We therefore explore a quantum dot CZ gate \cite{Meunier2011}; see Supplementary Information for theoretical considerations on the CZ operation in exchange based systems. This approach allows individual control over the qubits in the absence of interaction and its associated noise, while it uses the coupling to perform two-qubit operations with a frequency that can be much higher than the qubit Rabi rotation frequency, as we will now demonstrate.

As described by Eq.1 and depicted in Fig. \ref{fig:2}a, changing the detuning energy $\epsilon$ modifies the qubit resonance frequencies of $\textrm{Q}_1$ and $\textrm{Q}_2$ and introduces an effective detuning frequency $\nu_{\uparrow\downarrow,(\downarrow\uparrow)}$, such that one qubit acquires a time-integrated phase shift $\phi_{\uparrow\downarrow,(\downarrow\uparrow)}$, see Fig. \ref{fig:3}a, that depends on the $\sigma_z$-component of the other qubit, and vice versa. The exchange coupling and detuning frequency $\nu_{\uparrow\downarrow,(\downarrow\uparrow)}$ is maximised at the anticrossing $\epsilon=U$, see Fig. \ref{fig:2}a. When the exchange is non-zero, SWAP oscillations can also occur, but these are negligible as long as $\delta E_Z$ significantly exceeds the effective coupling $t$. When a CZ operation is performed such that $\phi_{\uparrow\downarrow}-\phi_{\downarrow\uparrow}=\pi$, the operation differs only by an overall phase from the basis CZ gate \cite{Ghosh2013}. This overall phase can be removed using single-qubit pulses or via voltage pulses exploiting the Stark shift {\cite{Veldhorst2014}. To realise a CNOT operation using the CZ gate, a CZ($\pi$) rotation is performed in between two $\pi/2$-pulses on $\textrm{Q}_2$ that have a phase difference $\phi_{\uparrow\downarrow}$. In Fig. \ref{fig:3}a we show the protocol of a CNOT gate for the case that $\phi_{\uparrow\downarrow}$ and $\phi_{\downarrow\uparrow}$ are equal and opposite, which is when $\delta E_Z \approx \sqrt{2}t_0$, where conveniently $\phi_{\uparrow\downarrow}=\pi/2$.

Figure \ref{fig:3}b shows the spin-up fraction $f_\uparrow$ of $\textrm{Q}_2$ after applying a $[\pi/2]_X$-pulse and a $[\pi/2]_Y$-pulse on $\textrm{Q}_2$ separated by an interaction time $\tau_Z$, with qubit $\textrm{Q}_1$ rotated to either $|\color{darkgreen}{\downarrow} \color{black} \rangle$ or $|\color{purple}{\uparrow} \color{black}\rangle$. The resulting exchange oscillations are initially a little slower due to filtering on the pulse line, but after \texttildelow 300 ns the oscillations approach a constant frequency of $\nu_{\uparrow\downarrow,(\downarrow\uparrow)}$  $\approx$ 2.4 MHz. The visibility with $\textrm{Q}_1$ $|{\uparrow}\rangle$ is slightly less due to preparation errors of $\textrm{Q}_1$. We have fitted the data and we find that the two sequences are out-of-phase ($\phi_{\downarrow\uparrow}$ = -$\phi_{\uparrow\downarrow}$), such that a CNOT gate occurs at $\tau_Z$ = 130 ns, when $\phi_{\uparrow\downarrow}=\pi/2$.

In Fig. \ref{fig:3}c we show experiments with different magnitudes of the exchange coupling, set via the detuning $\epsilon$. Plotting the frequency $\nu_{\uparrow\downarrow}$ as a function of $V_{P2}$, see Fig. \ref{fig:3}d, gives a trend consistent with that observed via ESR mapping, as shown in Figs. \ref{fig:2}b and c. The two-qubit coherence time $T^*_{2,CZ}$ is the free induction decay time of the two-qubit system. At large values of detuning ($\epsilon \rightarrow \infty$) or when the interaction vanishes ($t_0 \rightarrow 0$), the two-qubit coherence time $T^*_{2,CZ}$ simply reduces to the single-qubit Ramsey $T_2^*$. We obtain $T^*_{2,CZ}$ by fitting an exponential to the decay of the oscillations in Fig. \ref{fig:3}c. These values of $T^*_{2,CZ}$ are plotted along with the measured $\nu_{\uparrow\downarrow}$ values in Fig. \ref{fig:3}d. We find that the two-qubit dephasing rate $1/T^*_{2,CZ}$ rises in step with the exchange coupling and $\nu_{\uparrow\downarrow}$, which is to be expected since $\delta \nu_{\uparrow\downarrow}/\delta V_{P2}$ also increases with $\nu_{\uparrow\downarrow}$, meaning that the qubit system becomes increasingly sensitive to electrical noise. Despite this, the total number $N_{CZ}$ of two-qubit operations that can be performed also increases with $\nu_{\uparrow\downarrow}$, as shown in Fig. \ref{fig:3}d. In the Supplementary Information we show an optimized sequence where $T_2^*$ = 8.3 \textmu s and $\nu_{\uparrow\downarrow}$ = 3.14 MHz, wich gives the number of two-qubit CZ($\pi$) gates, $N_{CZ}=2\times$$(\nu_{\uparrow\downarrow}+\nu_{\downarrow\uparrow})$$\times$$T^*_{2,CZ}$$>$ 100. This indicates that the error can be less than 1\%, corresponding to a fidelity above 99\% for the two-qubit CZ gate. The fast two-qubit operation frequency implies also that over $10^5$ CZ gates can be performed within the single-qubit coherence time \cite{Veldhorst2014}. Future experiments will move towards devices with reservoirs on each side of the double quantum dot, so that both qubit states can be measured independently, opening the possibillity of full two-qubit tomography and a more detailed assessment of the fidelities. 

The tremendeous progress of quantum error correction codes over the last decade has resulted in schemes that allow fault-tolerant quantum computing with single and two-qubit errors as high as 1\% \cite{Fowler2012}; values that already seem consistent with the fidelities of these silicon quantum dot qubits. These qubit fidelities could be further improved by lowering the sensitivity to electrical noise. This could be achieved by designing the two-qubit system such that it is completely decoupled from the reservoir during qubit control, possibly by additional pulsing on the barrier gates. Additionally, increasing the inter-dot tunnel coupling will result in a lower $\delta \nu_{\uparrow\downarrow,(\downarrow\uparrow)}/\delta V_{P2}$ at the same $\nu_{\uparrow\downarrow,(\downarrow\uparrow)}$, while pulsing simultaneously on qubit top gates G1 and G2 would allow a larger on/off ratio of $\nu_{\uparrow\downarrow,(\downarrow\uparrow)}$, so that even faster CZ operations could be performed. Although these silicon qubits represent the smallest scalable two-qubit system reported to date, the complete fabrication process is remarkably compatible with standard CMOS technology, and is also consistent with current transistor feature sizes, offering the exciting prospect of realising a large-scale quantum processor using the same silicon manufacturing technologies that have enabled the current information age.


Acknowledgments: The authors thank S. Bartlett for useful discussions. The authors acknowledge support from the Australian Research Council (CE11E0096), the US Army Research Office (W911NF-13-1-0024) and the NSW Node of the Australian National Fabrication Facility. M.V. acknowledges support from the Netherlands Organization for Scientiﬁc Research (NWO) through a Rubicon Grant. The work at Keio has been supported in part by the Grant-in-Aid for Scientific Research by MEXT, in part by NanoQuine, in part by FIRST, and in part by JSPS Core-to-Core Program.

Author contributions:
M.V., C.H.Y. and J.C.C.H. performed the experiments. M.V. and F.E.H. fabricated the devices. K.M.I. prepared and supplied the 28Si epilayer wafer. W.H., J.P.D., J.T.M., S.S and A.L. contributed to the preparation of the experiments. M.V. C.H.Y. A.M. and A.S.D designed the experiment and discussed the results. M.V. analysed the results. M.V. and A.S.D. wrote the manuscript with input from all co-authors.

The authors declare no competing financial interests.

\newpage

Correspondence should be addressed to: 

M.V. (M.Veldhorst@unsw.edu.au) or 

A.S.D.(A.Dzurak@unsw.edu.au).

\clearpage
\begin{widetext}

\begin{center}
\textrm{\textbf{Supplementary Information}}

\textrm{\textbf{A Two Qubit Logic Gate in Silicon}}

\end{center}

\textbf{Experimental methods} \\
The device is fabricated on an epitaxially grown, isotopically purified $^{28}$Si epilayer with a residual $^{29}$Si concentration of 800 ppm \cite{Fukatsu2003}. Using a multi-level gate-stack silicon MOS technology \cite{Angus2007}, three layers of Al-gates are fabricated with a thickness of 25, 50 and 80 nm, separated by thermally grown oxidized aluminum on top of a SiO2 dielectric with a thickness of 5.9 nm. The measurements were conducted in a dilution refrigerator with base temperature $T_{bath}$=50 mK. DC voltages were applied using battery-powered voltages sources and added through resistive voltage dividers/combiners to voltage pulses using an arbitrary waveform generator (LeCroy ArbStudio 1104). Filters were included for slow and fast lines (10Hz to 80MHz). ESR pulses were delivered by an Agilent E8257D microwave analog signal generator and a 3 dBm attenuator at the 4 K plate. The stability diagrams are obtained using a double lock-in technique (Stanford Research Systems SR830) with dynamic voltage compensation.
	
  All our qubit statistics are based on counting the spin states of the quantum dot. Each data point represents the average of 200 up to 1000 single shot read outs, taken in 5 to 10 sweeps to compensate for small drifts. 
	Further details on single shot read out can be found in Veldhorst $et$ $al$ \cite{Veldhorst2014}.
\\
\\ 
\textbf{1. Device structure} \\
Figure 1a of the main text shows an SEM-image of the device, consisting of a charge sensor, an electron reservoir, an ESR line for qubit control and a quantum dot structure to define the qubits. We use an SET (pink) for charge sensing that is only capacitively coupled to the quantum dot structure (top-gates and barriers gates G1-4 are in blue, the confinement gate C is in brown). The qubits can be loaded and read using the reservoir R (brown) and controlled via the ESR-line (green).
\\
\\

\begin{figure*} [b!]
	\centering 
		\includegraphics[width=0.5\textwidth]{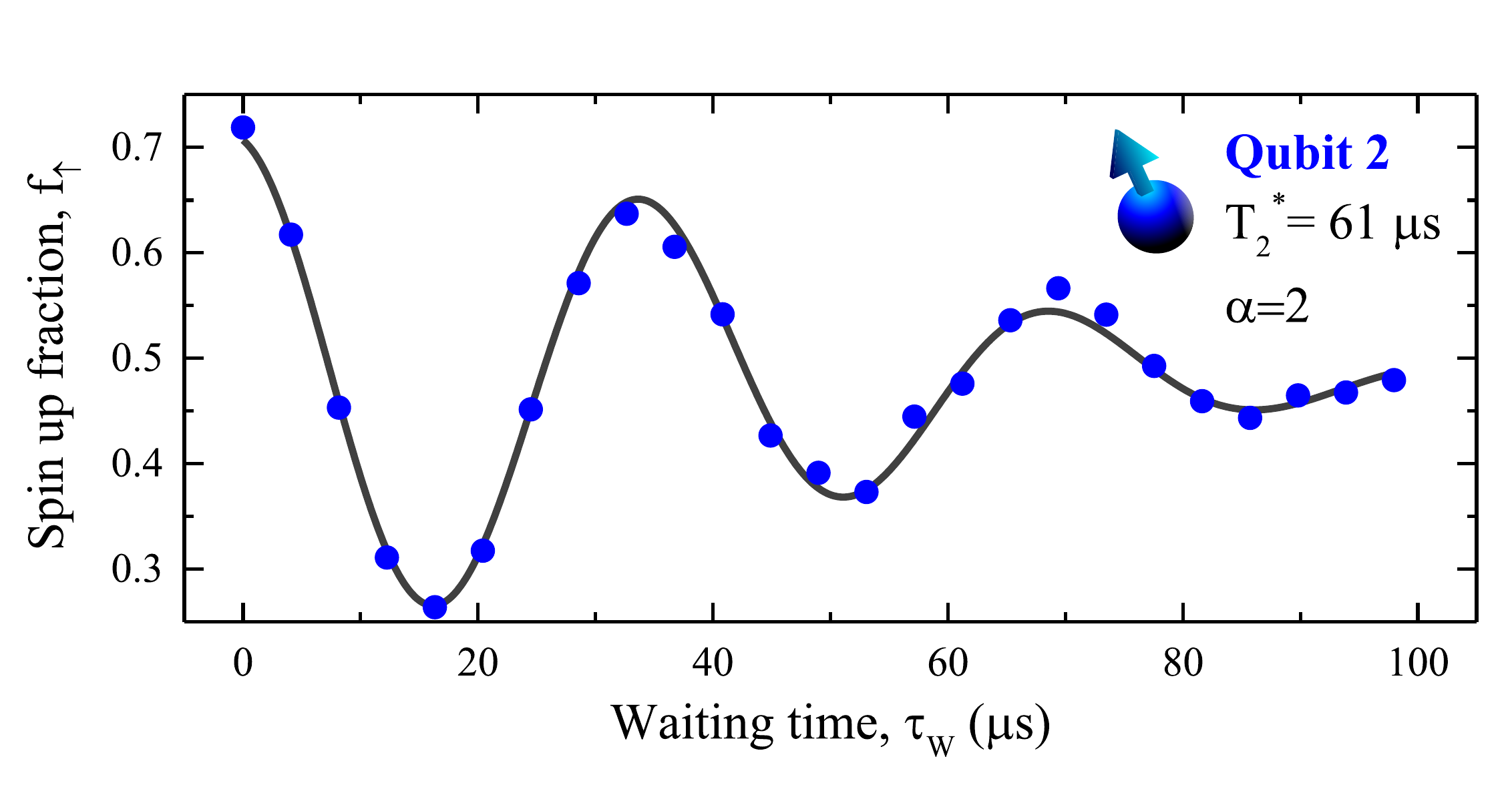}
		\caption{ \textbf{Ramsey fringes of qubit Q2.} A Ramsey experiment is performed, where two $\pi/2$ pulses are applied with increasing waiting time in between. From the decay we infer a $T_2^*$ = 61 \textmu s. The factor $\alpha=2$ is the power of the exponent and consistent with the power measured on Q1 \cite{Veldhorst2014}.}  
		\label{fig:S1}
\end{figure*}

\textbf{2. Single qubit coherence times} \\
Details of the coherence times of qubit Q1 can be found in Veldhorst $et$ $al$ \cite{Veldhorst2014}. Figure \ref{fig:S1} shows Ramsey fringes of qubit Q2, obtained by applying two $\pi/2$-pulses separated by a waiting time $\tau_w$. We have fitted the data with $f_\uparrow=A+B\cos(2\pi\delta_D \tau_w+\phi)\exp(-[\tau_w/T_2^*]^2)$, where $A$ and $B$ correct for measurement errors, $\delta_D$ is the detuning frequency, $\tau_w$ the waiting time, $\phi$ a phase offset and $T_2^*$ the dephasing time. With this, we find $T_2^*$ = 61 \textmu s.
\\
\\

\begin{figure*} []
	\centering 
		\includegraphics[width=0.5\textwidth]{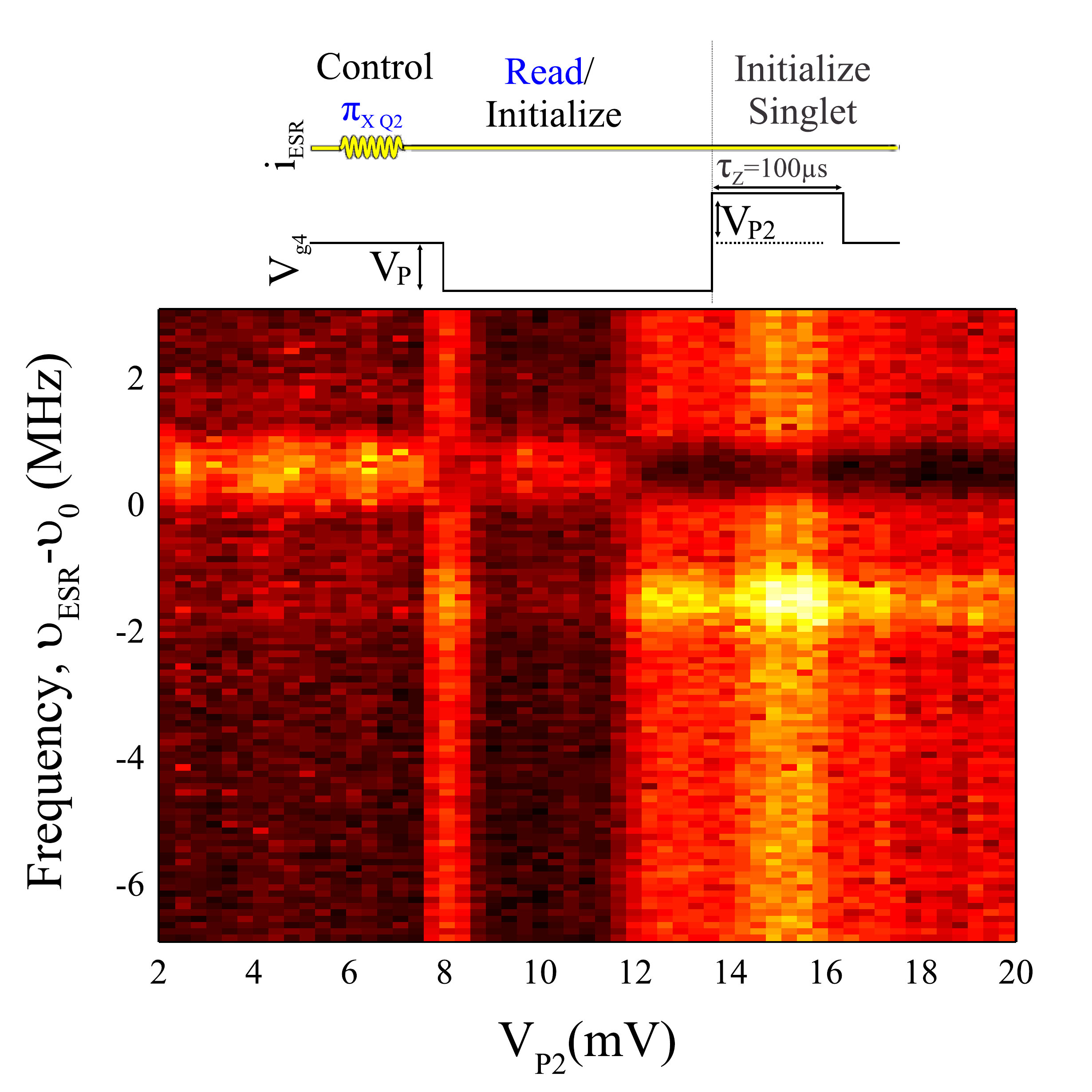}
		\caption{ \textbf{ESR control and voltage pulsing.} The ESR resonance frequency of Q2 depends on the state of Q1 due to the exchange coupling. Voltage pulsing towards the $\left|1,2 \right\rangle$ state results in singlet initialization. Another state is observed before the $\left|1,2 \right\rangle$ is reached, which might be an excited orbital or valley state.}  
		\label{fig:S2}
\end{figure*}

\textbf{3. Qubit initialization} \\
The qubits can be initialized to the $\left|\downarrow,\downarrow \right\rangle$-state by aligning the chemical potential of the reservoir in between the Zeeman splitted levels. Using microwave pulsing it is possible to rotate each qubit to the desired state. However, as shown in Fig. 2c of the main text we can also exploit the $\left|1,2 \right\rangle$-state, see Fig. 1e of the maint text, by applying a $V_{P2}$ to initialize the two-qubit system into a singlet state. In Fig. \ref{fig:S2}, we show the spin up fraction after applying a $\pi$-pulse on Q2 with different initialization (changing $V_{P2}$) and in the presence of finite interaction. For small $V_{P2}$ the qubit remains in the $\left|\downarrow,\downarrow \right\rangle$-state and the spin up fraction is zero, except when on resonance, where it is fully rotated. Note that there is in principle only one resonance, as Q1 is always $\left|{\downarrow} \right\rangle|$. At large $V_{P2}$, the charge state is (1,2). At the end of the $V_{P2}$-pulse, the electron on N2 tunnels to the reservoir and one of the two electrons on N1 tunnels to N2, such that Q1 and Q2 form a singlet. Applying a $\pi$-pulse on Q2 results then in a spin up fraction $f_{\uparrow}$= 1/2, except when on one of the two resonances, where it is rotated to either zero or one. Interestingly, we observe before the $\left|1,2 \right\rangle$-state another crossing ($V_{P2}$ = 8 mV), where the spin up fraction is similar as pulsing to the $\left|1,2 \right\rangle$-state. This level crossing with the $\left|\downarrow,\downarrow \right\rangle$-state is possibly due to an excited orbital or valley state. 
\\
\\

\begin{figure*} []
	\centering 
		\includegraphics[width=0.5\textwidth]{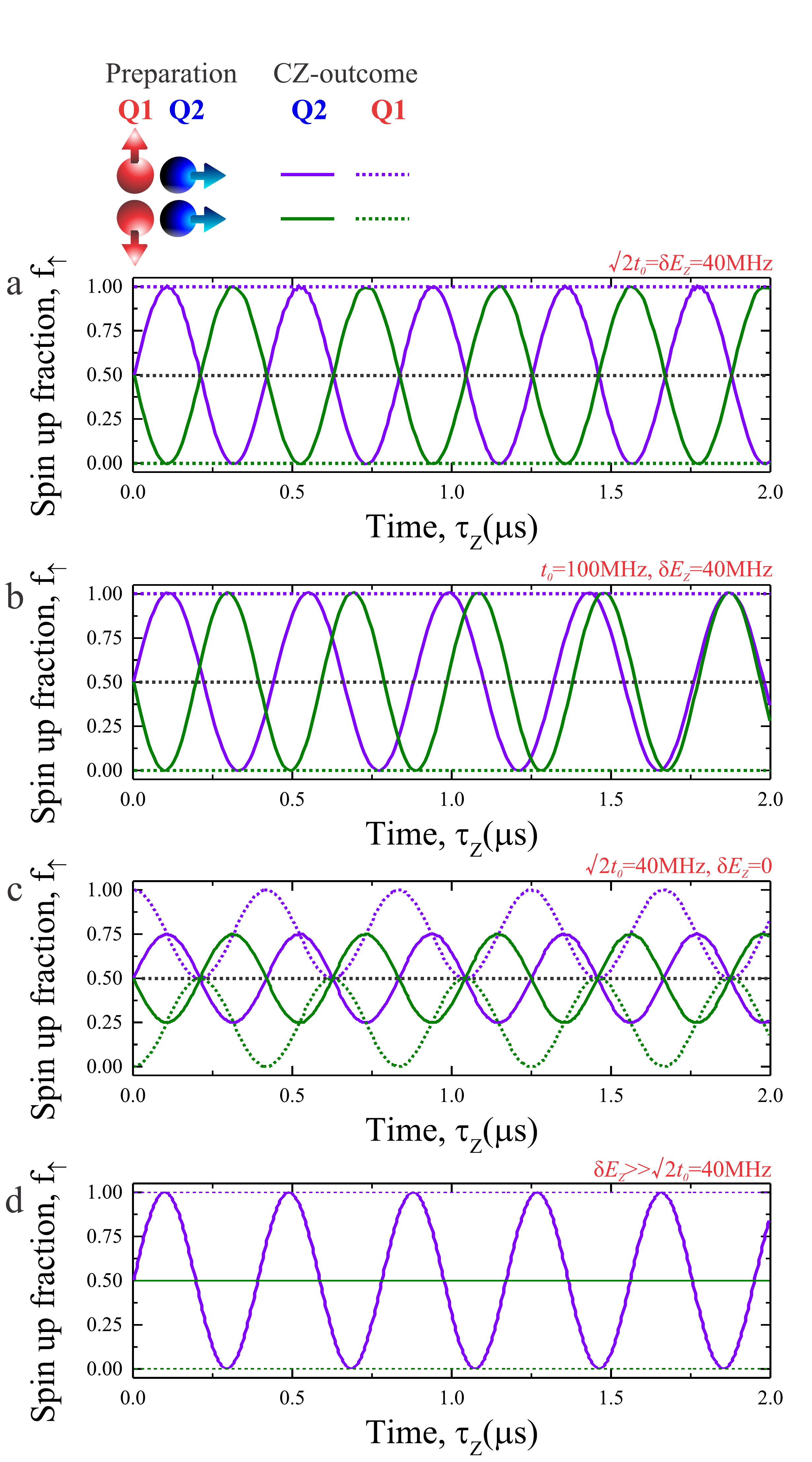}
		\caption{ \textbf{CPHASE using CZ.} Spin up fraction of the CZ operation as function of the interaction time $\tau_Z$ for different values of the Zeeman energy and tunnel coupling. When $\sqrt(2)t_0=\delta E_Z$ $\textbf{(a)}$ the CZ oscillations of the two $\sigma_z$ states are approximately equal and opposite; whereas for other values $\textbf{(b)}$ the oscillations evolve with different frequency. When the Zeeman energy difference is small, SWAP oscillations appear $\textbf{(c)}$ and when the Zeeman energy difference is very large $\textbf{(d)}$, rotations occur for only one $\sigma_Z$. A high-fidelity CNOT-gate using CZ can be constructed in principle in all parameter regimes by choosing the right single qubit operations.}     
		\label{fig:S3}
\end{figure*}

\textbf{3. controlled phase (CZ) operations} \\
Single qubit operations are performed at far detuning, eliminating the noise associated with the exchange coupling. While in principle two qubit operations are possible using ESR-pulses in the presence of exchange coupling, the coupling itself can be used directly to allow fast two-qubit operations. The effect of the exchange coupling strongly depends on the relative values of the Zeeman energies $E_{Z1}$ and $E_{Z2}$ and the tunnel coupling $t_0$. In silicon there is a degeneracy in valley, which can have its impact on the operation as well. However, in a regime where the valley splitting $E_V \gg t_0$, only one valley is relevant. Also, MOS double quantum dots have Coulomb repulsions with on-site energies $U \gg t_0$, such that coupling between the $\left|0,2 \right\rangle$-states and the $\left|2,0 \right\rangle$-states are vanishingly small. Therefore, the system is well described using only single-particle terms (excepth for the on-site Coulomb interaction), resulting in Eq.1 of the main text.

Figure \ref{fig:S4} shows the time evolution of the CZ-operations in different parameter regimes. Here, two Rabi $\pi/2$ pulses with a phase difference $\phi$ = $\pi/2$ performed at far detuning on the target qubit are separated by a waiting time $\tau_Z$ in the presence of interaction between the two qubits. When $\sqrt{2}t_0=\delta E_Z$, the in-plane oscillations of the target qubit with the control qubit initialized $\left|\uparrow(\downarrow) \right\rangle$ are nicely out-of-phase (Fig. \ref{fig:S3}a), producing a high fidelity CNOT operation when $\tau_Z$ $\approx$ 100 ns. For other ratios of $t_0/\delta E_z$ (Fig. \ref{fig:S3}b), high fidelity operations can still be obtained by correcting the frequency difference using a different phase $\phi$ = $\phi_{\uparrow\downarrow}$, see Fig. 3 of the main text, between the two Rabi $\pi/2$ pulses. When the exchange coupling becomes too large with respect to the Zeeman energy difference, the qubits start to make SWAP operations (Fig. \ref{fig:S3}c), which have to be corrected with more complex pulses. In the extreme that $\delta E_Z \gg t_0$ (Fig. \ref{fig:S3}d), in-plane oscillations occur for only one $\sigma_Z$.
\\
\\

\begin{figure*} []
	\centering 
		\includegraphics[width=0.8\textwidth]{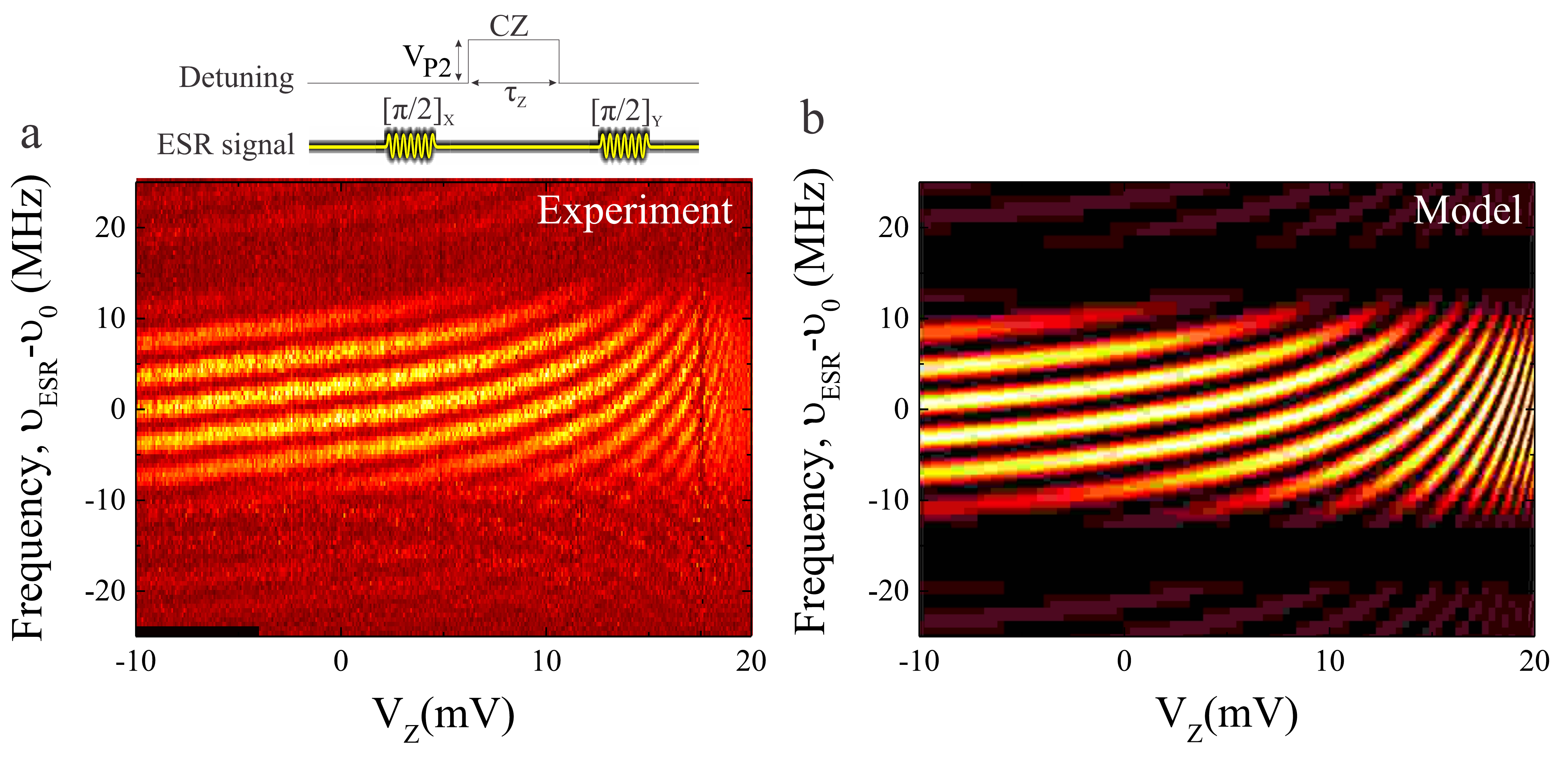}
		\caption{ \textbf{XZY-sequence.} Resulting spin up fraction after two Rabi pulses with length $\pi/2$ separated with a $V_{P2}$-pulse with length $\tau_Z$ = 1.7 \textmu s and changing the height of the pulse. Increasing the voltage pulse, increases the exchange coupling, resulting in in-plane oscillations. The experiment \textbf{a} nicely coincides with a model \textbf{b} based on the $\left|1,1 \right\rangle$ to $\left|0,2 \right\rangle$-coupling. In the vertical direction, the slow oscillations correspond to a Rabi pulse with length $\pi$, the faster oscillations correspond to in-plane Ramsey rotations due to finite detuning frequency. Moving along the horizontal direction changes the frequency of the in-plane rotations due to the exchange coupling. Due to imperfect initialization in the experiment, Q1 is sometimes in the state $\left|\uparrow \right\rangle$, resulting in the (less visible) downward bending lines.}     
		\label{fig:S4}
\end{figure*}

\begin{figure*} [h!]
	\centering 
		\includegraphics[width=0.9\textwidth]{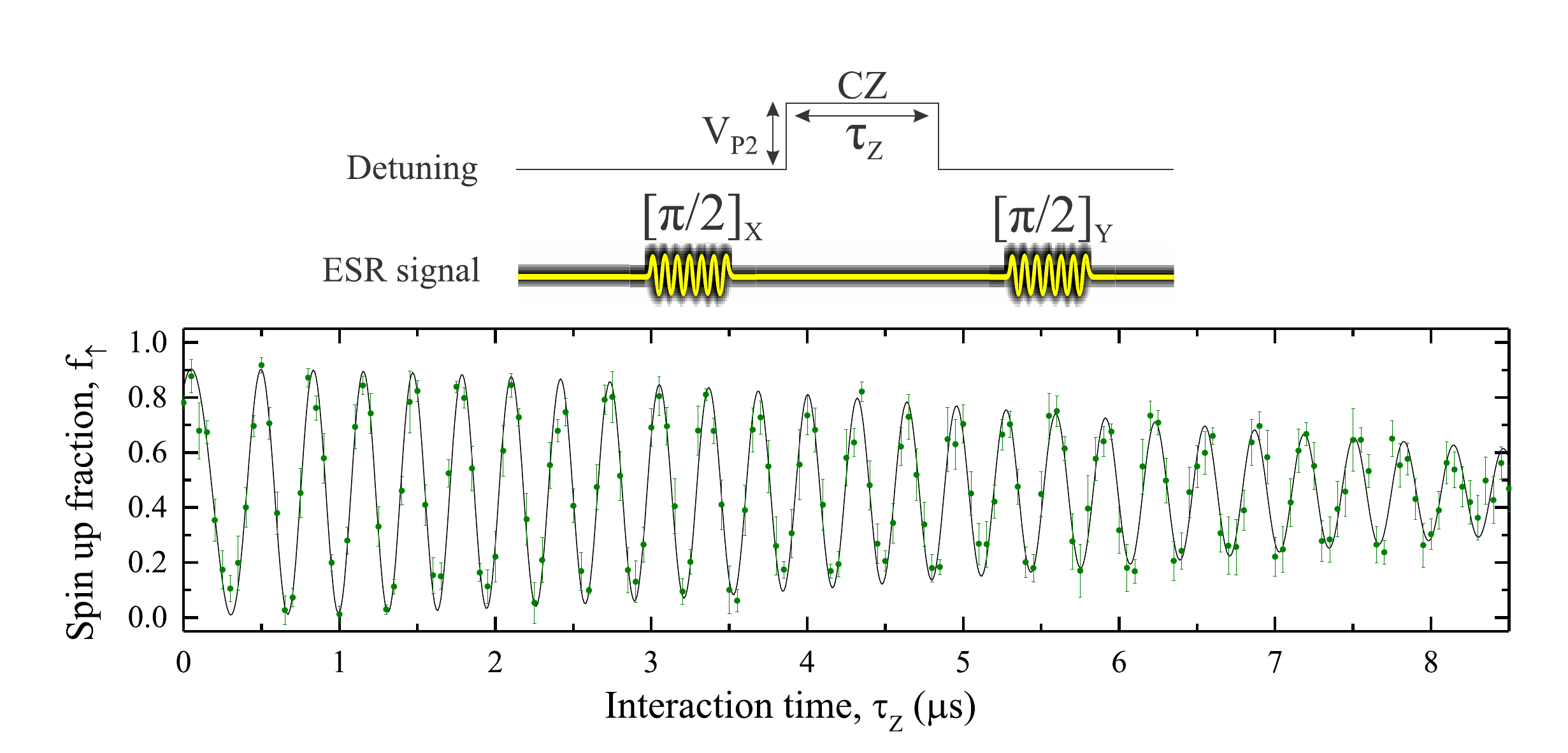}
		\caption{ \textbf{Two qubit CZ-operation.} Spin up fraction after two Rabi rotations with length $\pi/2$ and phase difference $\phi=\pi/2$ are separated with increasing time $\tau_Z$. The presence of exchange coupling results in CZ-oscillations with frequency $\nu_{\uparrow\downarrow}=3.14$ MHz and a decay constant $T_{CZ}$ = 8.3 \textmu s. The errors bars correspond to the standard error of the mean of 210 single shot events per data point.}  
		\label{fig:S5}
\end{figure*}

\newpage

\textbf{4. XZY-control and CZ-operations} \\
We have explored the CZ-performance using an XZY-pulsing sequence, consisting of an X-pulse and Y-pulse of length $\pi/2$ and calibrated to have phase difference $\phi=\pi/2$. The time $\tau_Z$ is fixed to 1.7 \textmu s, equal to a Rabi $\pi$-rotation. The result is shown in Fig. \ref{fig:S3}, where the 'upward bending' of the lines are due to an increase of the in-plane rotation frequency (determined by the detuning frequency and the exchange coupling).

After carefully tuning the qubit, we are able to observe over 25 CZ-oscillations, see Fig. \ref{fig:S5}. The two-qubit system has a coherence time $T_{CZ}$ = 8.3 \textmu s at this exchange coupling resulting in $\nu_{\uparrow\downarrow}$ = 3.14 MHz, which corresponds to over 100 CZ($\pi$)-operations within the coherence time.
\\
\\

\end{widetext}

\end{document}